# Controlled thinning and surface smoothening of Silicon nanopillars

[1]S. Kalem*, [2]P. Werner, [3]B. Nilsson, [2]V.G. Talalaev, [3]M. Hagberg, [3]Ö. Arthursson, [3]U. Södervall

[1]TUBITAK – UEKAE, The Scientific and Technical Research Council of Turkey – National Institute of Electronics and Cryptology, Gebze 41470 Kocaeli, TURKEY
2Max-Planck-Institute of Microstructure Physics, Weinberg 2, D-06120 Halle, Germany
3Department of Microtechnology and Nanosciences, Chalmers University of Technology, Göteborg, Sweden

E-mail: s.kalem@uekae.tubitak.gov.tr

**Abstract:**

A convenient method has been developed to thin electron beam fabricated Silicon nanopillars under controlled surface manipulation by transforming the surface of the pillars to an oxide shell layer followed by the growth of sacrificial ammonium silicon fluoride coating. The results show the formation of an oxide shell and a Silicon core without significantly changing the original length and shape of the pillars. The oxide shell layer thickness can be controlled from few nanometers up to few hundred nanometers. While down sizing in diameter, smooth Si pillar surfaces of less than 10 nm roughness within 2 μm were produced after exposure to vapors of HF and $HNO_3$ mixture as evidenced by transmission electron microscopy (TEM) analysis. The attempt to expose for long durations leads to the growth of a thick oxide whose strain effect on pillars can be assessed by coupled LO-TO vibrational modes of Si-O bonds. Photoluminescence (PL) on the pillar structures which have been down sized exhibit visible and infrared emissions, which are attributable to microscopic pillars and to the confinement of excited carriers in Si core, respectively. The formation of a smooth core-shell structures while reducing the diameter of the Si pillars has a potential in fabricating nanoscale electronic devices and functional components.

# 1. Introduction

Faster, efficient and cost-effective processes for a further size reduction of semiconductor nanostructures (by top-down approach) to create smooth (defect free) functional surfaces and interfaces are of crucial importance for nanotechnology applications [1-3]. There are number of advantages of thinning semiconductor pillars, wires or rods from few hundred nanometers of diameter down to desired sizes. First of all, most of the useful quantum mechanical confinement effects take place at reduced dimensions well under the Bohr radius [4]. For example, the effect of size on the electronic band structure in Silicon starts from about 30 nm whose strength depends on crystal orientation and effective mass at low dimensions[5, 6]. Further interesting phenomena can take place at lower dimensions. For Silicon nanowires (SiNWs), it has been shown that bulk symmetry is not conserved due to reduced transverse dimensions increasing the transverse mass due to quantum confinement and Si becomes a direct band gap semiconductor for <100>, <111> and <110> orientations [7]. As a consequence of this modification, very efficient optical transitions, for example photoluminescence from thinned Si pillars can be expected from the band-edge transitions at the zone center. In addition, subsequent enlargement of the band gap is naturally expected due to size effects thus offering tunable optical properties. Recent tight binding calculations taking into account true effective masses (instead of 3D effective mass approximation) predict related changes in the band structure of Silicon pillars in both conduction and valence bands and its effects on electronic device properties[7-9].

In addition to quantum size effects, there are several other factors affecting the electronic band structure of Si pillars or wires. It has been shown that the band structure of SiNWs can be modified by strain [10] and type of surface termination [11]. In the case of <100> and <111> oriented NWs, tensile strain enhances the direct band gap characteristics, whereas compressive strain attenuates it. Surface termination significantly modifies the band gap, resulting in relative red energy shifts in competition with quantum confinement [11]. On the other hand, tapering induces also changes on electronic band structure of SiNWs by creating a strong potential gradient along the wire axis [12].

From the applications point of view, top-down approach of pillar production offers more precise control over size and shape as compared to bottom-up methods [13]. Massive pillars with much better surface quality can be grown by bottom-up methods, however the need for a metal seeding, alignment, location control, silicidation process still introduce certain disadvantages from electronic device fabrication point of view [14, 15]. While the size and the shape control is an advantage for the top-down approach, down-sizing in diameter and the presence of electronic surface defects due to plasma etching are among major concerns [16, 17]. Therefore, there is a need of a novel size-reduction method for the production of wires, rods or pillars having desired dimensions and surface quality which are, e.g., functional for electronic device fabrication.

In fabricating high aspect ratio nanopillars, inductively coupled reactive ion etching techniques have been proven to be very effective [17, 18]. However, due to the inherent nature of chlorine ($Cl_2$) reactive ion etching of crystalline Silicon, the surfaces around the pillars come out as rough. Similar surface roughness has also been created during electrochemical size reduction of Silicon nanowires [19-21]. The roughness is usually accompanied by undesirable surface defects which are detrimental for optical and electronic device applications. Therefore, the fabrication of nanopillars with defect free smooth surfaces is required for implementation to semiconductor manufacturing. Our work is addressing this crucial issue, and it is proposing a method for the smoothening of plasma etched nanopillars while reducing the diameter of Si pillars with a controlled manipulation. The thinning of Silicon pillars in such a controlled way can be considered as a step forward for the production of functional semiconductor surfaces for electronic and photonic applications.

## 2. Experimental

The Si nanopillars studied here were fabricated by using electron beam lithography which was followed by inductively coupled plasma/reactive ion etching (ICP/RIE) using chlorine ($Cl_2$) plasmas. This process used a chlorine flow rate of 50 sccm corresponding to a process pressure of 7mTorr under an electrode power and inductively coupled power of 50 Watts and 100 Watts, respectively. The pillars with targeted diameters of either 250 nm or 500 nm were prepared on a surface area of 5mmx5mm with a lattice constant (**a**) of i) **a**=500 nm and ii) **a**=1000 nm, respectively. The scanning electron microscope (SEM) image of such as-fabricated Si pillar cluster is shown in Fig. 1. Note that due to RIE etching, the pillars have a conical shape having about 20% change in diameter from the tip to bottom. The caps on the pillars are residual plasma etch protection oxide. For imaging, the pillars were tilted and these pillars were the same pillars as those used in the following TEM images (see Fig.4 and Fig.5) taken after processing (acid vapor exposure).

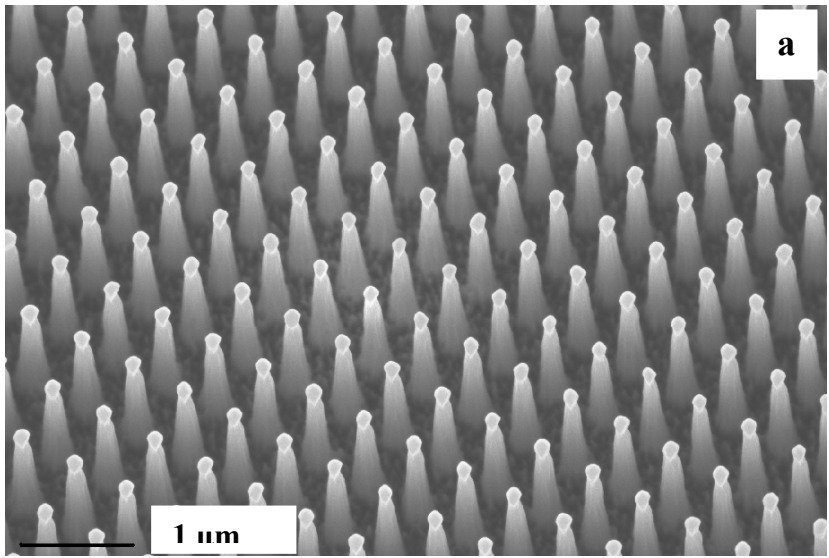

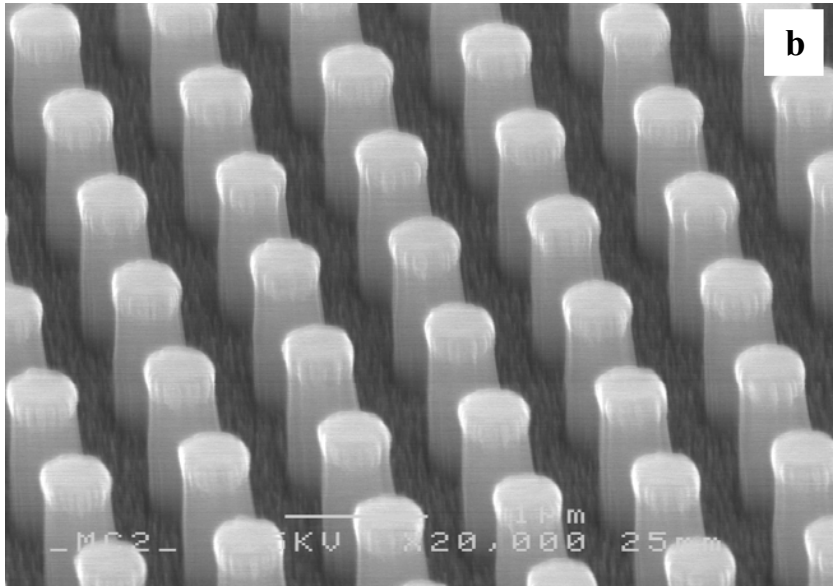

**Figure 1.** SEM image of as-fabricated Si pillar clusters before acid vapor exposure. (a) 250 nm. (b) 500 nm. The caps on the pillars are residual etch protection oxide and the background consists of microscopic pillars or black Si.

The thinning of Si nanopillars, rods, wires down to desired sizes has been realized by a controlled manipulation of surface structures via a room temperature oxide growth and etching by exposing the surface of the wafer to vapors of HF and $HNO_3$ acid mixture. Before e-beam patterning, a thermal oxide layer of about 1000 nm was grown on both <100> and <111> oriented wafers for plasma etch protection. The oxide growth on the sidewalls of the nanopillars or size reduction of Si pillars was realized by exposing the surface of pillars to vapor of $HF:HNO_3:H_2O$ (70:25:15) acid mixture. The chemicals used in the process were semiconductor grade %48 HF and %65 $HNO_3$ by weight. Further details of this experiment were described elsewhere [22-24]. The etch protection oxide as a cap layer can be removed from the tips of the pillars by a diluted HF dip or it can also be removed during the exposure to the acid vapor. As a parameter characterizing the process, we present the change in thinning or diameter reduction as measured at the half-width of the maxiumum height for each pillar as shown in Fig. 2. This evolution can be approximated by an exponential growth or two slope linear change corresponding to a slow and fast mechanism with the thinning rates of 4.8±1.5% and 47±15% per minute, respectively. Also, before the sample characterization including photoluminescence, TEM and Fourier transformed infrared spectroscopy (FTIR) measurements, any residual ammonium silicon hexafluoride $(NH_4)_2SiF_6$ (hereafter ASH) has been removed from the pillars by rinsing the sample for a few seconds in de-ionized (DI) water (see N-H modes in Section **3**). Photoluminescence has been excited by a HeCd laser operating at 325 nm. The resulting emission was detected by an Acton Spectrophotometer using CCD, InGaAs array and Ge photodetector at room temperature as well as at 10K. FTIR measurements were done at room temperature between 400 – 4000 $cm^{-1}$. SEM and TEM images were recorded using a JEOL JSM 6340F and a JEOL JEM 4010, respectively.

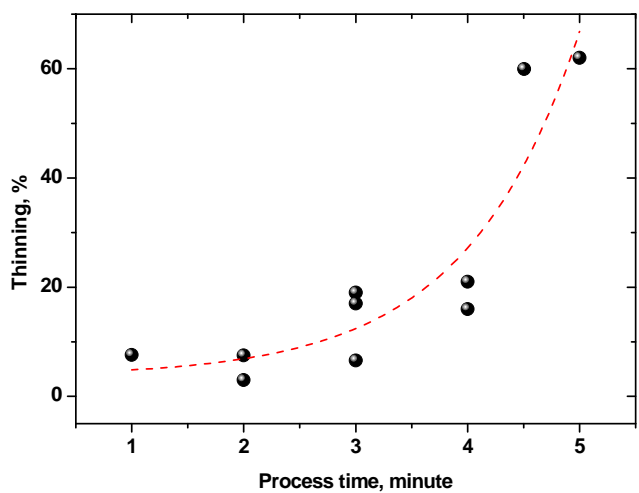

**Figure 2.** Thinning rate or diameter reduction of Si pillars versus acid vapor exposure time.

## 3. Results and discussion

The sketch in Fig. 3 describes a typical structural detail of a Si nanopillar array after it has been exposed to vapors of HF:HNO$_3$ acid mixture. Following this process the surface of the pillars are transformed to ASH [22-24] and a thin oxide layer interface as a shell S was formed between the ASH and Si pillar (core) C. The ASH can be readily removed by an H$_2$O rinse after the process. Note also that the etch protection oxide (E) at the top of the pillar can be removed during the exposure to acid vapors.

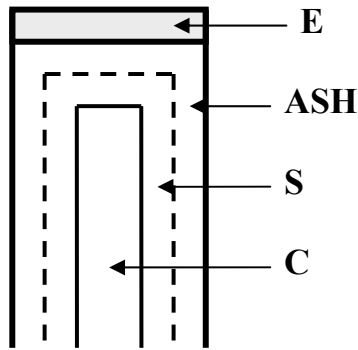

**Figure 3.** A sketch describing the structure of a pillar after it has been exposed to vapors of HF:HNO$_3$ mixture. The letters correspond to plasma etch protection oxide (E), sacrificial layer of ammonium silicon hexafluoride (ASH), Silicon oxide shell (S) and a Silicon core (C).

Figure 4a shows the TEM cross-sectional image of acid vapor treated Silicon nanopillars of about 4 micrometer long which were fabricated on <111> oriented Si wafers. These pillars have then been exposed to a vapor of HF:HNO$_3$ acid mixture for a duration of 1 minute followed by DI water rinse to remove any residual ASH. The vapor exposure produces a Silicon oxide layer around the pillars whose thickness depends on the exposure time. The oxide layer around the pillars could also be removed in a diluted (5%) HF dip after this process. Note that the nanopillars are not wiped out during the exposure to the harsh acid vapor environment. Instead, the pillars are smoothly thinned with sharp sidewalls as shown in

Fig. 4a. In this particular set of pillars, the reduced diameter of pillars amounts to around 175 nm but it gets thinner toward the tip corresponding to a size reduction of about 10%. However, some of the pillars were found to be overetched as shown in Fig. 4b particularly from the tip parts resulting in a needle-like feature following the original shape of the as-fabricated pillars (see Fig. 1). Thus, the pillars have been thinned down to 175 nm from 200 nm (pillar width at half-height) by the transformation of the pillar surfaces to an intermediate oxide layer followed by the growth of a sacrificial layer of ASH [22-24], which could be removed by rinsing the pillars in water. We observe that the core of Si nanopillars is still crystalline as evidenced by Bragg reflections (dark regions in Fig. 4a and 4b) and also evidence is provided by diffraction patterns indicating a <111> pillar orientation (see Fig. 4c). Despite the exposure to strong acids, the nanopillars are not wiped out, instead smooth surfaces were obtained for pillars which have been exposed to vapors of $HF:HNO_3$. A closer examination of the TEM picture confirms the presence of very smooth pillar surfaces ( see Fig. 4d, where the arrow points at the surface). Figure 4e shows a TEM image of the pillar surface over longer scale. From these experiments we estimate a surface roughness of about less than 10 nm within 2 μm. In corresponding high-resolution TEM image, ordered atomic planes and single crystalline structure can be observed in the nanopillars. Spacing of lattice fringes of 3.14 Å between interatomic planes perpandicular to wafer plane agrees reasonably well with that between Si atoms on Si (111) wafers treated by HF [25] and Silicon nanowires with their axes lying along the [111] direction [26]. Fig. 4f shows the statistical distribution of pillar diameter before and after processing. About 5-10% of change in diameter can be achieved for an exposure time of 60 second. The process can be further improved by working with diluted vapors.

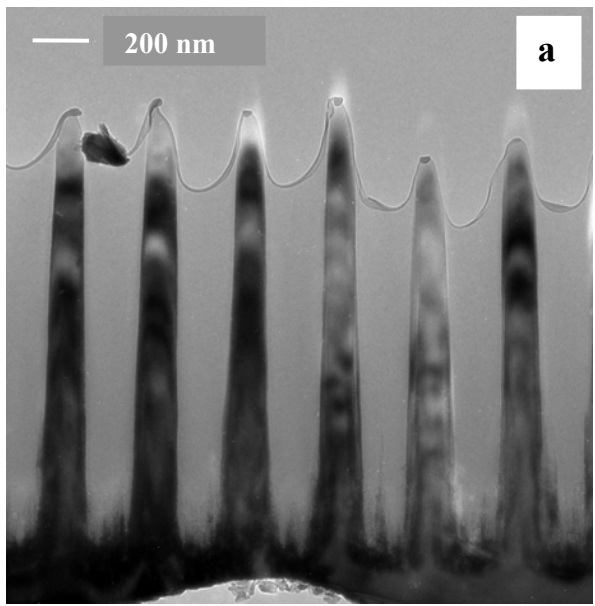
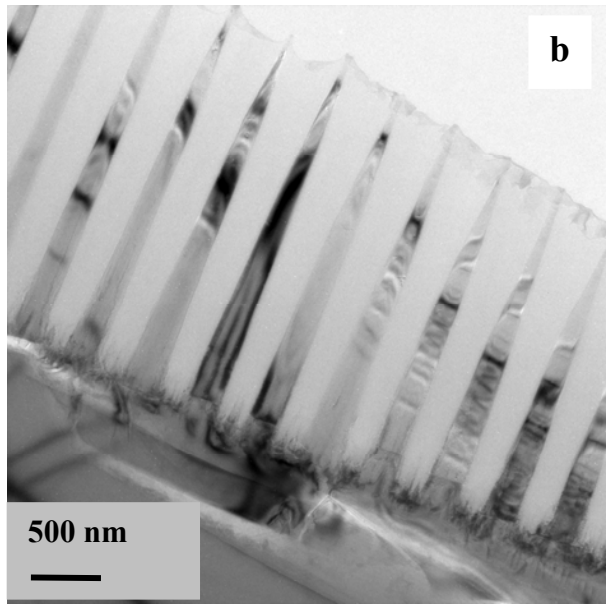
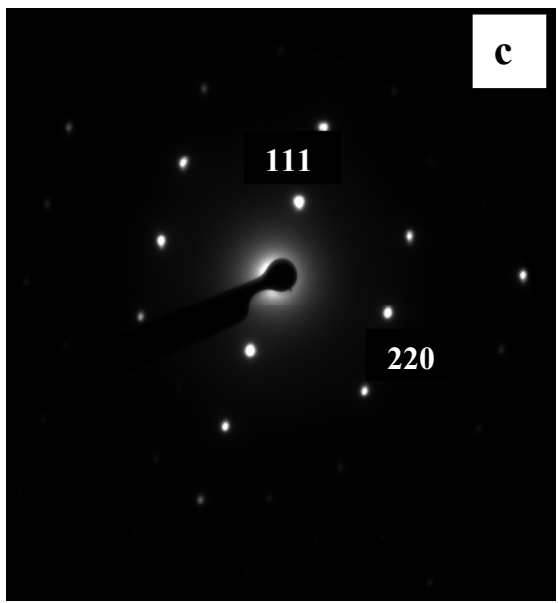
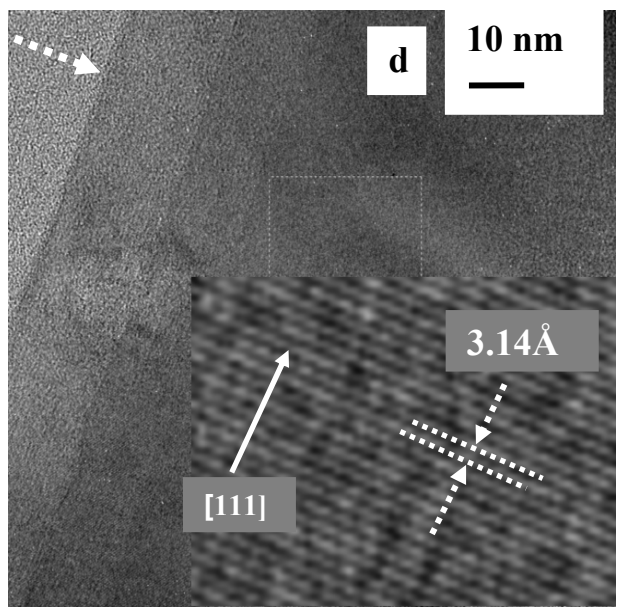

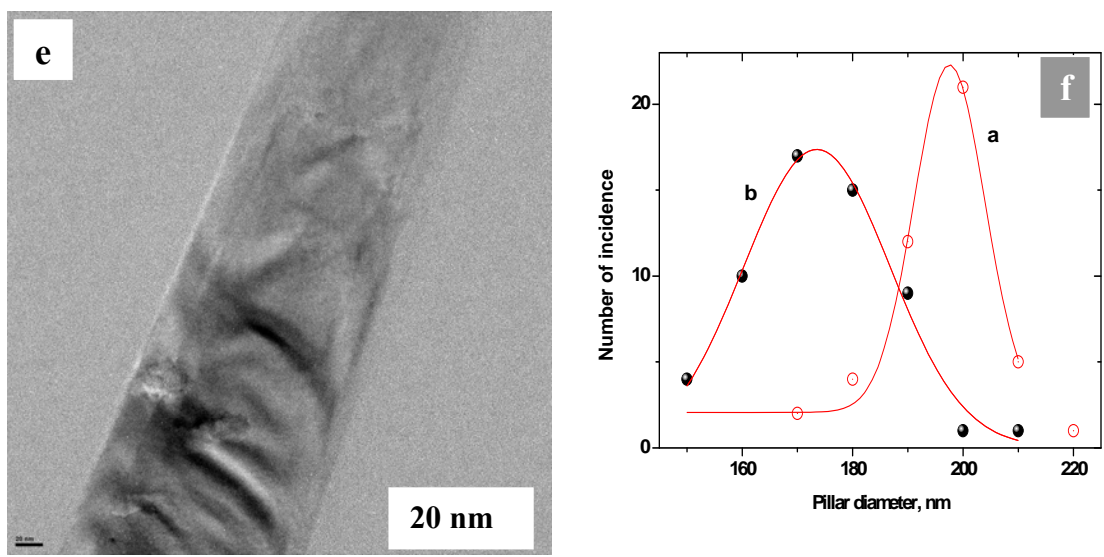

**Figure 4.** TEM analysis of thinned pillars. (a) Cross-sectional image of thinned e-beam fabricated pillars for 1 minute exposure. (b) The effect of over exposure resulting in needle like pillars. (c) Electron diffraction pattern confirming the crystalline structure of the nanopillars. (d) The image of e-beam lithography fabricated pillars showing the formation of smooth surface of Si pillars. The arrow indicates the surface and the insert shows the interference fringes of atom planes. (e) TEM image of pillars showing the surface smoothness over larger scale. f) Histogram showing the diameter distribution: a) before processing, b) after processing.

For excessive etching times corresponding to a thick oxide growth on the pillar sidewalls, there might be a lattice-strain build-up. In the following particular example of Fig. 5a showing the cross-sectional SEM image, the outer part of Si pillars of 500 nm was transformed to a thick oxide layer of 150 nm as a shell using the same vapor exposure method. As it was mentioned in the previous section, the roughness around the pillars is due to RIE etching within the chlorine plasma. Note that our method produces an homogeneous oxide layer without compromising significantly from the original length of the nanopillars that is about 1.5 μm. A Si core with rather a weak contrast was observed in the TEM image of the same pillars (see Fig. 5b). However, cross-sectional gray-scale profiling clearly indicates the presence of the core-shell structure of the pillars (see Fig. 5c). The two minimums in this figure shows the limits between the Si core of 165 nm and the Si oxide shell layer. Note that during this process no damage was caused to pillars and also the etch protection oxide layer was smoothly removed from the top of the pillars as shown in the Fig. 5b. The TEM image reveals the presence of a needle shaped Silicon core pointing toward the top of the pillars. The weak contrast between the core and the oxide shell is probably significative of the presence of a compositional gradient of the oxide. Moreover, the formation of the thick oxide layer causes a $\Lambda$ shaped kink with an angle of almost 90º at the bottom of each pillar (Fig. 5b). This kink is the result of the homogeneous oxide growth ( of more than 100 nm) which spans over all the surface of the wafer and the pillars.

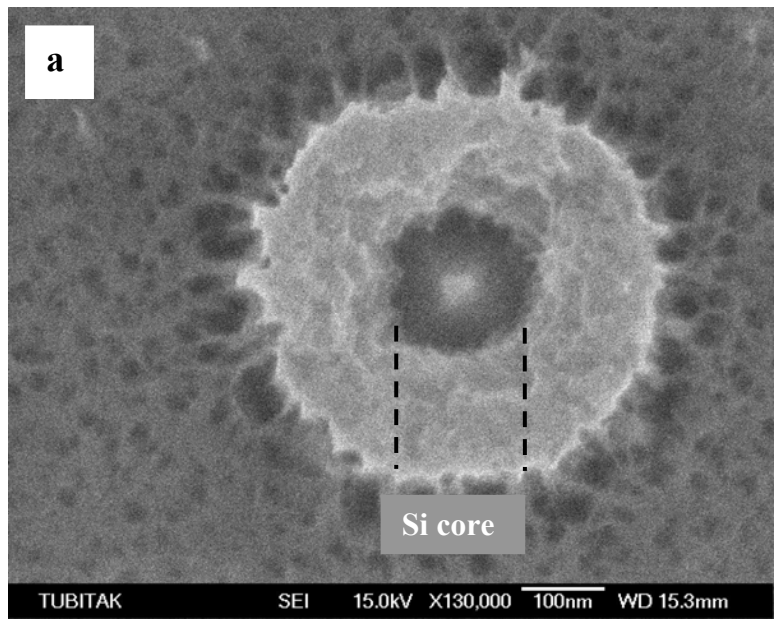

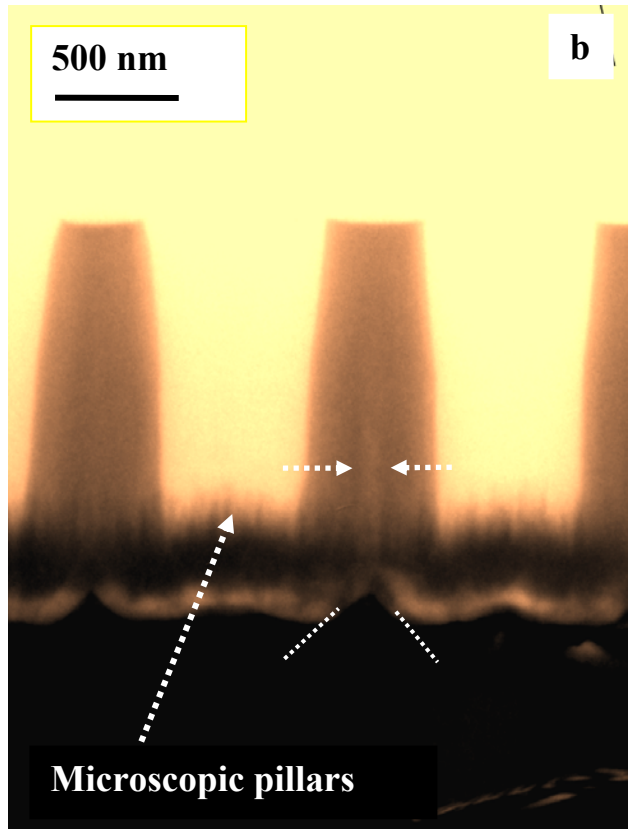

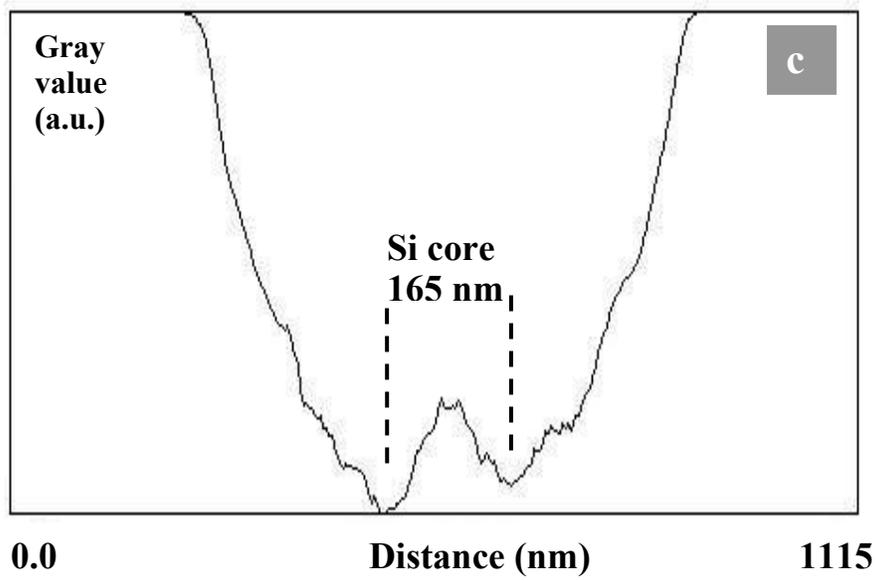

**Figure 5.** (a) Cross-sectional SEM viewgraph of an intentionally broken pillar indicating the presence of a Si core of 165 nm and an oxide shell of 150 nm. (b)TEM image of thick oxide

shell/Si core structure (marked by arrows) for pillars processed for 4 minutes. Note that there is a Λ shaped kink formation (dashed lines) between each pillar and the Si wafer which is the result of a contrast formed by the same oxide. Microscopic pillars between the pillars also indicated by a dashed arrow, c) gray-scale profiling of the TEM image indicating the limits (two minimums) between the Si core and oxide shell.

In order to confirm the presence and the evolution of possible strain in the pillars, oblique-incidence dependence of the absorbance of vibrational modes were measured in 1000-1300 cm$^{-1}$ region. The FTIR spectra of such vapor-phase oxidized Silicon pillars are shown in Fig. 6 as compared to the as-fabricated one and the one just before DI water rinse (spectrum b). Note that the major N-H (nitrogen-hydrogen) modes (3300 cm$^{-1}$, 1430 cm$^{-1}$ and 735 cm$^{-1}$) relating $(NH_4)SiF_6$ species in spectrum b are disappeared after DI water rinse in spectrum c. The pillar before and after acid treatment, exhibits LO-TO coupled frequency pairs at 1080cm$^{-1}$ and 1255cm$^{-1}$, but the total oxygen amount (as it can be deduced from the integrated intensity of the oxygen peak at 1080 cm$^{-1}$) is much larger when the sample is exposed to acid vapors. The LO-TO splitting mode pair could be an indicative hint to the presence of stress, disorder or compositional changes in the film [27, 28]. When relative strengths of bands at 1088 cm$^{-1}$ and 1255 cm$^{-1}$ are compared, the ratio of integrated intensities $I_{1088}/I_{1255}$ decreases from 1.29 to 1.20 thus confirming the relative enhancement of the splitting mode at 1255cm$^{-1}$ by 7.5% for the vapor processed pillars. The enhanced oscillator strength of the mode at 1255cm$^{-1}$ as the incidence angle is changed shows that there is likely a presence of strain or disorder build-up in the pillar structure due to this thick oxide layer [28]. Right now, it is not possible to determine what would be the amount of the contributions of stress, disorder or composition change to 7.5% of increase. Further systematic studies taking into account the role of micropillars are required to find out the contributions from each possible source. In addition to these observations, the peaks at 2115 cm$^{-1}$ and 620 cm$^{-1}$ in FTIR spectra are Si-H stretching and wagging related modes, respectiveley, indicating also the presence of bonded hydrogen in the oxide layer.

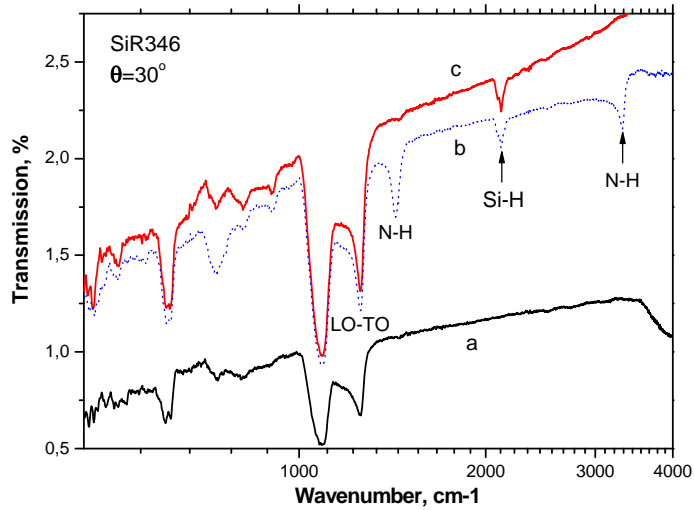

**Figure 6.** FTIR spectrum indicating strain related Si-O LO-TO coupling modes at 1080 cm-1 and 1255 cm-1. Also, Si-H vibration at 2115 cm-1 is an evidence for the incorporation of hydrogen. The figure compares the spectrum of as-fabricated pillars(a), pillars after vapor exposure process confirming the presence of an ASH layer through N-H vibrations (b) and pillars after DI water rinse(c).

Room temperature (RT) photoluminescence (PL) properties are summarized in Fig. 7 indicating strong emissions in the visible but also in the near infrared region when the Silicon pillars are excited by a laser line of 325 nm. In the visible part of the spectrum, strong RT Gaussian shaped emission ranging from 550 nm to 850 nm and peaking at around 690 nm (1.80 eV) with a band width of 175 nm. We could observe the emission even by naked eye. The enhancement of light is partly the result of photonic confinement of the light emission due to the photonic structure (lattice constant a=1000 nm) of the pillars. The origin of the emission itself is not sufficiently clear although it could be attributed to porous Silicon like structure containing small Si clusters passivated by an oxide layer [29]. The presence of hydrogen (H) bonded to Silicon as observed in FTIR spectra confirms that H atoms are likely passivating dangling bonds of related defects, which are detrimental to radiative recombination. However, at low temperatures, a broader and stronger emission band was observed exhibiting several peaks as shown in Fig. 7a at 560 nm (2.21 eV), 605 nm (2.05 eV), 645 nm (1.92 eV) and 705 nm (1.76 eV). Investigations on a number of samples combined with a Gaussian curve deconvolution prove that the emission bands at 605 nm, 645 nm and 705 nm exist also at room temperature, but their intensity becomes much stronger at 10 K. However, the band at 560 nm was emerged as a new band and only observed at low temperature. The inset in Fig 7a shows the PL spectrum of microscopic pillars lying at the bottom of the main pillars (see Fig. 5b). We observe a strong emission at 560 nm suggesting that the PL emission in pillar structures originates from these microscopic pillars. At present, it is not clear whether the emission is due to quantum confinement or oxide related defects. If the quantum confinement is in effect, these observations suggest that the band at 560 nm is due to Si nanostructures having sizes of less than 10 nm or 1 nm if a cross-sectional area of 1 nm was assumed [30]. Then the enhancement of the PL intensity can be attributed to more efficient carrier capture within these nanostructures. The microscopic pillars, also known as black Silicon [30], are the great source of light emission contributing to the photoluminescence at 560 nm.

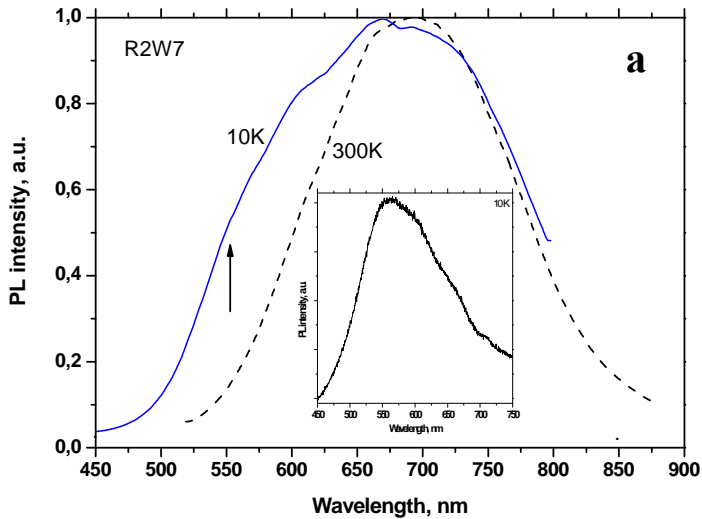

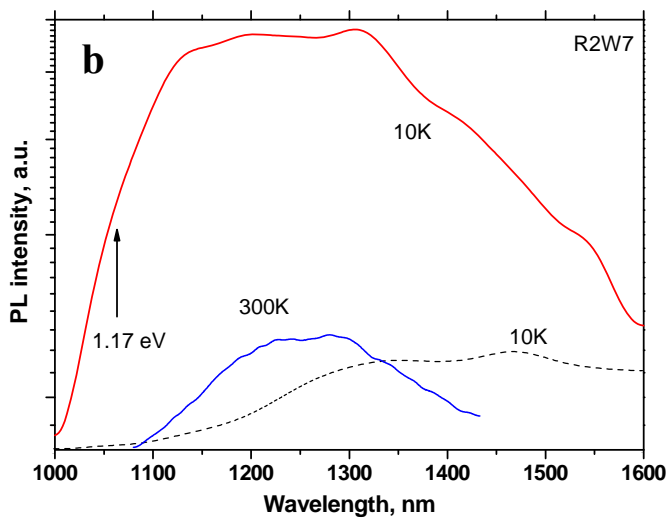

**Figure 7.** (a) VIS photoluminescence of thinned e-beam fabricated pillars at room temperature and 10K. The inset is the PL spectrum of background microscopic pillars at 10K. (b) Band-edge photoluminescence of thinned Si pillars at RT and 10K. Also shown is the band-edge emission from the microscopic pillars (dashed line).

Beside the visible emission band, we observe also a significant RT photoluminescence at around 1250 nm, which has been associated with Silicon band-edge emission as shown in Fig.7. In bulk Silicon, there is no photoluminescence at room temperature due to indirect nature of the band gap, that is very low transition probability. However, recent investigations have shown that the bulk crystal symmetry is not preserved in the NWs due to the nanometer size of the transverse dimensions, thus enabling strong optical transitions at the Brillouin zone center [31, 32]. This suggests that phonon assistance is not required for optical transitions since the maximum of the conduction and valence bands are aligned at **k=0.** The effect of reduced dimensions on the band structure of Si nanowires has been described elsewhere in greater detail [6, 7]. Low temperature PL spectra were recorded to support the RT data. These results indicate that the band-edge PL should be attributed to recombination within the Si pillars. Transfer of electron-hole pairs from the surrounding oxide layer to Si pillars is the most likely mechanism for the enhanced band-edge PL as it has been suggested for Si crystallites embedded in $SiO_2$ [33]. At low temperature, the emission is stronger and similar to the one at RT, with the exception of a tail at high energies. The Gaussian deconvolution analysis of the emission reveals the position of this tail at 1063 nm (1.17 eV) that is most likely caused by non-phonon (NP) line emission corresponding to the band gap energy of Silicon at 10K. These PL results don't support the presence of any quantum size effect. This is partly due to the fact that the pillar's sizes are still far from the quantum mechanical regime wherein the pillar diameter **R** should be less than 30 nm. However, some of the pillars possess a needle like structure with tip dimensions of less than 10 nm as a result of over etching (Fig. 4b). But, from the fact that there is no blue shifted PL emission, we suggest that electron-hole pair (excitons) created within the tips are fast diffusing out of this region before a radiative recombination can occur. Exciton diffusivities of 11 $cm^2/s$ at 12 K in Silicon have been reported [34] to occur by capture-release mechanism through neutral impurities. This effect should be further enhanced by tapering induced potential gradient [12]. Thus, the excitons created within this region (R < 30 nm) are diffused or transferred toward low energy region (R > 30 nm) where they are recombined radiatively leading to observed band-edge emission as described in Fig. 8.

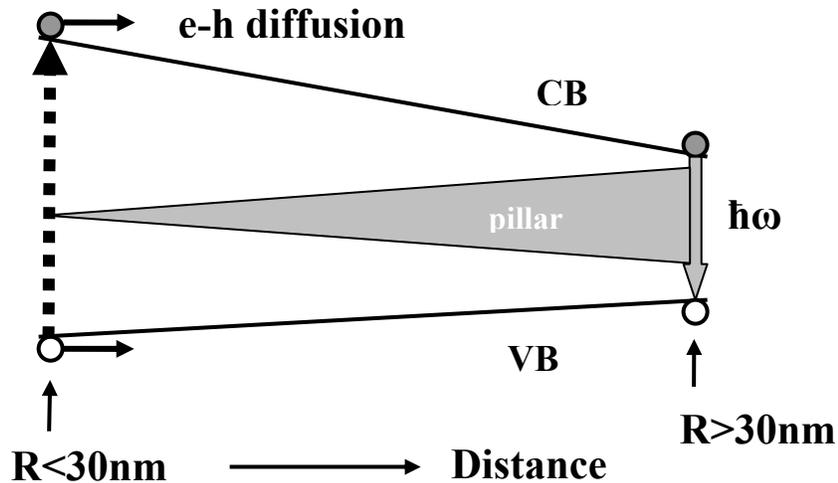

**Figure 8.** A sketch illustrating the creation and diffusion of excitons from the possible confinement region toward the non-confinement region.

Low temperature PL from microscopic Si pillars is also shown in Figure 7b. The emission is located at relatively lower energies, that is less than 1.0 eV, confirming that observed transitions from our samples should be originated from the Si pillars.

## 4. Conclusion

We have shown that e-beam lithographically fabricated Silicon nanopillars can be thinned down to desired small dimensions in a controlled way without compromising from the original length of the pillars by exposing pillars to vapors of acid mixtures consisting of HF and $HNO_3$. By this method, the nanopillars were transformed to a core-shell structure with an oxide shell grown at room temperature around the pillars with a growth rate of up to 0.5 nm/sec. We show that the Si core is still crystalline with a smooth oxide shell surfaces having a roughness of less than 10 nm within 2 μm. Longer exposure times ( >3 minutes) leads to faster growth resulting in thick oxide shell (>150 nm) around the 500 nm diameter pillars. We show that coupled LO-TO modes of Si-O bonds can be used to assess the magnitude of the strain caused by the oxide shell. Photoluminescence spectroscopy experiments exhibit emissions in both visible and infrared regions which are originated from background microscopic pillars and confinement (not quantum size effect) of excited carriers in down-sized core of Silicon pillars, respectively. A controlled reduction of the pillar diameter and the formation of smooth Si core-oxide shell pillar structures can be applied to the fabrication of functional nanoscale silicon electronic devices.

**Acknowledgement:**

This work was supported by BMBF-TÜBITAK programs (contract No: TBAG-U/232-107T624).

**References**

[1] Stern E, Klemic J F, Routenberg D A, Wyrembak P N, Turner-Evans D B, Hamilton A D, LaVan D A, Fahmy T M, and Reed M A 2007 Nature **445** 519

[2] Chan C K, Peng H, Liu G, McIlwroth K, Zhang X F, Huggins R A, Cui Y 2008 Nature Nanotechnology **3** 31

[3] Zheng G, Patolsky F, Cui Y, Wang W U, Lieber C M 2005 Nature Biotechnology **23** 1294

[4] Lockwood D J and Pinczuk A 2002 Optical Phenomena in Semiconductor Structures of Reduced dimensions, Springer, NY

[5] Scheel H, Reich S and Thomson C 2005 Phys. Stat. Sol.(b) **242** 2474

[6] Nehari K, Cavassilas N, Autran J L, Bescond M, Munteanu D, Lannoo M 2006 Solid State Electronics 50 716

[7] Harris C and O'Reilly E P 2006 Physica **E32,** 341

[8] Trani F, Contele G, Ninno D, Iadonisi G 2005 Phys. Rev. **B72** 075423

[9] Nehari K, Cavassilas N, Michelini F, Bescond M, Autran J L, and Lannoo M 2007 Applied Physics Letters **90** 132112

[10] Hong K-H, Kim J, Lee S-H, and Shin J K 2008 Nanoletters **8** 1335

[11] Nolan M, O'Callaghan S, Fagas G, and Greer J.C. 2007 Nanoletters **7** 34

[12] Wu Z, Neaton J.B, and Grossman J.C. 2008 Physical Review Letters **100** 246804

[13] Pennelli G 2009 Microelectronic Engineering doi:10.1016/j.mee.2009.02.032

[14] Schubert L, Werner P, Zakharov N D, Gerth G, Kolb F M, Long L, Gösele U, Tan T Y 2004 Applied Physics Letters **84** 4968

[15] Liu B, Wang Y, Dilts S, Mayer T S, and Mohney S E 2007 *Nano Lett.* **7** 818

[16] Oehrlein G S, Rembetski J F and Payne E H 1990 J. Vac. Sci. Technol. **B 8** 1199

[17] Laws G M, Handugan A, Eschrich T, Boland P, Sinclair C, Myhajlenko S, Poweleit C D 2007 J.Vac.Sci.Technol.**B 25** 2059

[18] Chekurov N, Grigoras K, Peltonen A, Franssila S and Tittonen I 2009 Nanotechnology **20** 65307

[19] Fennandez-Serra M V, Adessi C, Blase X 2006 Nanoletters **6** 2674

[20] Juhasz R, Elfström N and Linnros J 2005 Nanoletters **5** 275

[21] Juhasz R and Linnros J 2001 Applied Physics Letters **78** 3118

[22] Kalem S and Yavuz O 2000 Optics Express **6** 7

[23]. Kalem S 2004 Applied Surface Science **236** 336

[24]. Kalem S 2008 Superlattices and Microstructures **44** 705

[25] Morita Y, Miki K, Tokumoto H 1991 Appl. Phys. Lett. **59**, 1347

[26] Ge S, Jiang K, Lu X, Chen Y, Wang R, and Fan S 2005 Advanced materials **17** 56

[27] Innocenzi P, Falcaro P, Grosso D and Babonneau F 2003 J. Phys. Chem. **B 107** 4711

[28] Kirk C T. 1988 Phys Rev **B 38** 1255

[29] Canham L T 1990 Appl. Phys. Lett. **57** 1046

[30] Sorokin P B, Avramov P V, Chernozatonskii L A, Fedorov D G, and Ovchinnikov S G, J. 2008 Phys. Chem. **A 112** 9955

[31] Zheng Y, Rivas C, Lake R, Alam K, Boykin T B and Klimeck G 2005 IEEE Trans. On Electron. Devices **52** 1097

[32] Gnani E, Reggiani S, Gnudi A, Parruccini P, Colle R, Rudan M and Baccaroni G 2007 IEEE Trans Electron Devives **54** 2243

[33] Tsybeskov L, Moore K L, Hall D.G, Faıchet P M 1996 Physical Review B **54** R8361

[34] Chen Y H and Lyon S A 1989 IEEE J. Of Quantum Electronics **25** 1053

**FIGURE CAPTIONS**

**Figure 1.** SEM image of as-fabricated Si pillar clusters before acid vapor exposure having targeted diameters of; (a) 250 nm. (b) 500 nm. The caps on the pillars are residual etch protection oxide and the background consists of microscopic pillars or black Si.

**Figure 2.** Thinning rate or diameter reduction of Si pillars versus acid vapor exposure time.

**Figure 3.** A sketch describing the structure of a pillar after it has been exposed to vapors of HF:$HNO_3$ mixture. The letters correspond to plasma etch protection oxide (E), sacrificial layer of ammonium silicon hexafluoride (ASH), Silicon oxide shell (S), Silicon core (C).

**Figure 4.** TEM analysis of thinned pillars. (a) Cross-sectional image of thinned e-beam fabricated pillars for 1 minute exposure. (b) The effect of over exposure resulting in needle like pillars. (c) Electron diffraction pattern confirming the crystalline structure of the nanopillars. (d) Image of e-beam lithography fabricated pillars showing the formation of smooth surface of Si pillars. The arrow indicates the surface and the insert shows the interference fringes of atom planes. (e) TEM image of pillars showing the surface smoothness over larger scale. f) Histogram showing the diameter distribution: a) before processing, b) after processing.

**Figure 5.** (a) Cross-sectional SEM viewgraph of an intentionally broken pillar indicating the presence of a Si core of 165 nm and an oxide shell of 150 nm. (b)TEM image of thick oxide shell/Si core structure (marked by arrows) for pillars processed for 4 minutes. Note that there is a $\Lambda$ shaped kink formation (dashed lines) between each pillar and the Si wafer which is the result of a contrast formed by the same oxide. Microscopic pillars between the pillars also indicated by a dashed arrow, c) gray-scale profiling of the TEM image indicating the limits (two minimums) between the Si core and oxide shell.

**Figure 6.** FTIR spectrum indicating strain related Si-O LO-TO coupling modes at 1080 cm-1 and 1255 cm-1. Also, Si-H vibration at 2115 cm-1 is an evidence for the incorporation of hydrogen. The figure compares the spectrum of as-fabricated pillars(a), pillars after vapor exposure process confirming the presence of an ASH layer through N-H vibrations (b) and pillars after DI water rinse(c).

**Figure 7.** (a) VIS photoluminescence of thinned e-beam fabricated pillars at room temperature and 10K. The inset is the PL spectrum of background microscopic pillars at 10K. (b) Band-edge photoluminescence of thinned Si pillars at RT and 10K. Also shown is the band-edge emission from the microscopic pillars (dashed line).

**Figure 8.** A sketch illustrating the creation and diffusion of excitons from the possible confinement region toward the non-confinement region.